\documentclass[amsmath,amssymb,reprint,bibnotes]{revtex4-2}
\usepackage{graphicx}
\usepackage{dcolumn}
\usepackage{bm}
\usepackage{amsmath} 
\usepackage[mathlines]{lineno}
\usepackage{braket}
\begin{document}
\title{Modulating the temporal dynamics of nonlinear ultrafast plasmon resonances}
\author{Hira Asif}
\author{Ramazan Sahin}
\affiliation{Faculty of Science, Department of Physics, Akdeniz University, 07058 Antalya, Turkey}
\date{\today}
\begin{abstract}
Spatio-temporal control of ultrafast plasmon resonances has gained research interest in recent years because of their tremendous implications in nonlinear optics and ultrafast quantum technology. In particular, the lifetime of ultrashort plasmon oscillations has become a debatable subject in recent experimental and theoretical studies to fulfill the future challenges concerning their effective employment in the vast applications of the plasmonic industry. Here, we examined the temporal properties of nonlinear plasmonic modes in metal nanostructures by interacting them with quantum objects in the weak coupling regime in order to distinguish it from the fundamental plasmonic mode. First of all, we present an analytical description of nonlinear ultrafast dynamics of localized surface plasmon resonances when the second harmonic plasmon mode interacts with long-lived dark mode or quantum emitter. Later, the coupled plasmonic system is realized in two different ways to control the lifetime of second harmonic mode by coupling, i) driven mode to dark mode (or long lifetime quantum emitter) ii)itself to dark mode (or long lifetime quantum emitter). The driven-dissipative dynamics are solved through a numerical technique governing the spatial and temporal changes in the second harmonic plasmonic response supported by AuNP. Finally, the lifetime enhancement of nonlinear plasmon mode is manifested by performing FDTD simulations for a nonlinear plasmonic system of Au nanoparticles coupled with a long lifetime quantum emitter.
\end{abstract}
\maketitle
\section{Introduction} 

In the light-matter interaction at the nanoscale, the conduction electrons of noble metal nanoparticles oscillate collectively with a resonant frequency known as localized surface plasmon resonance (LSPR). The frequency and the strength of the oscillation depend on the size, shape and the composition of the nanoparticles \cite{Giannini2011,bookchapter}.
This LSPR on the surface of metallic nanoparticles (MNP) exhibits exotic optical properties, \cite{Soukoulis2011} such as near field enhancement in the subwavelength regime and large optical cross-sections in the visible and near-infrared portions of the electromagnetic spectrum \cite{Barnes2006}. These plasmonic features of metallic nanoparticles lead to their use in the exciting applications such as solar cells \cite{Ueno2018}, nanoantennas \cite{Giannini2011}, colorimetric sensing \cite{Mirkin1996,Stewart2008,Anker2009}, nonlinear optical converters \cite{Panoiu2018}, single-molecule Surface-Enhanced Raman Scattering \cite{Kneipp1997} and plasmonic nano-cavity lasers \cite{Ma2011}.
Apart from strong spatial confinement and intense oscillatory field strength, the plasmonic field also possesses a predominant feature, i.e, the plasmonic field amplitude damps quickly within a few tens to hundreds of femtosecond \cite{Lazzarini2017, TFranz2002}. \\
The dephasing of the plasmonic field through fast electron scattering and nonradiative decay pathways has become a debatable subject in recent literature. For instance, the rapid decay rate of localized surface plasmon resonance (LSPR) in a metallic nanoparticle is attributed to radiative and nonradiative damping which primarily depends on the electron concentration in the size of nanoparticle \cite{Melikyan2004} and plasmon resonance frequency \cite{Scharte2001}. The investigation of the damping characteristics of surface plasmons for different nanostructures has been done through theoretical models such as quantum mechanical calculations of Landau damping  \cite{Kirakosyan2016} by using random phase approximation and density functional theory (DFT) \cite{Li2013, Yuan2008}. While quantifying the reason for the fast decaying plasmonic field, some theoretical works have also been done to elongate the period of plasmon oscillations. A long oscillation time of plasmonic yield has been suggested in \cite{M.Divece2018} through the large electric field enhancement in the nanogap region between two nanoparticles. These nanogaps attribute a quantum tunneling regime to accelerate high-energy electrons for increasing the efficiency of the electric field in plasmonic nanostructures. Decreasing the nanogap size between nanoparticles increases the intensity which in turn enhances the lifetime of the plasmonic field in contrast with the response of the field outside the nanogap. In the same way, the increase in electric field intensities of linear and nonlinear plasmon modes (but not lifetimes) have been suggested through Fano resonance \cite{Fanoresonance} in the steady-state while strong-coupling causes splitting in the plasmon mode \cite{sahin_graphene_splitting}. On the other hand, a recent study demonstrates that interacting nanostructures which support two different plasmon modes(bright and dark) can significantly alter the ultrafast response even in a very weak coupling regime \cite{Yildiz2020}. Increasing the lifetime of plasmon response by coupling two nanostructures offered many potential applications such as high-efficiency solar cells and enhanced Raman processes. For instance, coupling a longer lifetime quantum emitter lengthens the plasmon oscillation time of metal nanostructures enabling their use not only in optical nanoantennas and solar cell applications \cite{qelife2013} but also it enables imaging of individual molecules in aperture-less Scanning Near-field Optical Microscope (SNOM) \cite{sahin_single_molecule}.    
 
\begin{figure*}[hbt!]
\includegraphics[scale=0.5]{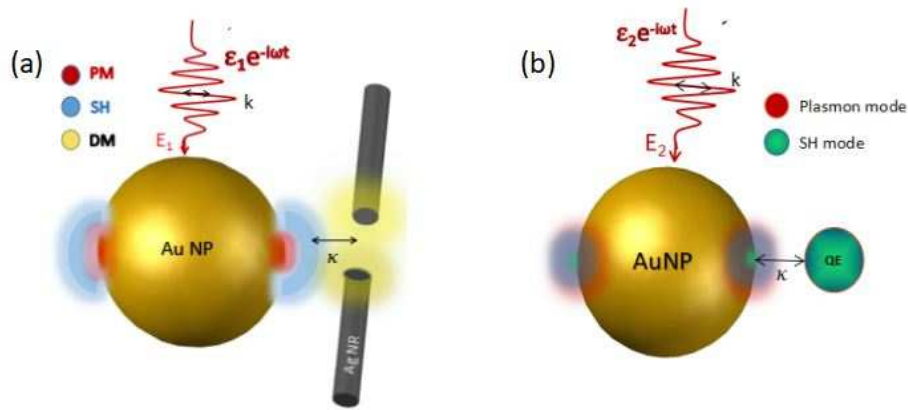}
\caption{Two coupled plasmonic systems of metallic nanostructures under the action of EM field E with polarization vector k in the x-direction (a) SH LSPR blue region around AuNP couples to dark mode of AgNR (b) second harmonic (SH) mode green region couples to QE.}
\label{fig1}
\end{figure*}

Here in this paper, we aim to couple short-lived plasmon oscillations to long-lived coherent modes, which is in our case supported by Ag nanorod (AgNR) and a two-level quantum emitter (QE) so that linear and non-linear plasmonic fields can be distinguishable via adjusting the lifetime of nonlinear plasmonic field as a proof-of-principle concept. In the experiments, this can be performed by measuring the electric field intensity with time-delay (nearly 20 fs-this duration is enough for decaying of fundamental plasmon field) to acquire pure nonlinear plasmon field. The nonlinear plasmon mode, i.e. second harmonic (SH) mode, optimizes through the hybridization of driven plasmon mode (PM), which is excited by the electromagnetic plane-wave at resonance frequency $\omega$ in an Au nanoparticle (AuNP). The AuNP plasmonic system undergoes a resonant coupling when it is brought near to AgNR supporting dark plasmon mode or a long life QE as shown in Fig.\ref{fig1} (a and b), respectively. The second harmonic signal in the near field regime upon weak interaction deactivates the dark mode that turns out the exchange of energy and an intense spatial localization between these nanostructures. Since the SH signal from AuNP with a short lifetime, couples to the resonant oscillating field of long-lived dark mode or QE,  it not only enhances the plasmon field strength effectively in the steady-state but also lengthens its decay time dramatically. In the next section, we design two phenomenological models for a plasmonic system of AuNP coupled to AgNR as a dark mode in one case and QE in the second case. In both cases, we solve numerically the nonlinear system of differential equations and drive the nonlinear field amplitude to analyze the time evolution of SH field response.

\section{Analytical Models}
The model systems consist of two interacting nanostructures analog to coupled harmonic oscillators with a weak coupling strength \cite{Yildiz2015, Fanoresonance}. The combined plasmonic systems have coherent plasmon resonances with modified lifetimes. These lifetimes are associated with the bright and dark plasmon modes (or QE) of the nanostructures manifesting underdamped oscillations. For instance, the bright plasmon mode has a large decay rate around $\gamma_{p}\approx10^{14}$ Hz as compare to the small decay channel of dark mode, that is $\gamma_{d}\approx10^{12}$ Hz \cite{Yildiz2020}.
Whereas the decay rate of the quantum emitter (or quantum dot) is of the order of $\gamma_{q}=10^9$ Hz, which also supports an oscillating system with a small decay time. The coupling of these oscillators in-cooperates by adjusting the distance between the nanoparticle and other nanostructures which are defined as parameter $\mathcal{K}$ for coupling of SH with AgNR or QE as shown in Fig.\ref{fig1} (a,b). 
\subsection{AuNP-AgNR Coupled Nonlinear Plasmonic System}
The coherent dynamics of the coupled system of AuNP and AgNR (supporting dark plasmon mode), is described by the Hamiltonian as given \cite{Fanoresonance},
\begin{eqnarray}
\hat{\mathcal{H}}=\sum_{j=1}^{3}\omega_j\hat{a_j}\textsuperscript{\textdagger}\hat{a_j}+\mathcal{K}(\hat{a_2}\textsuperscript{\textdagger}\hat{a_3}+\hat{a_3}\textsuperscript{\textdagger}\hat{a_2})
+\varepsilon_1\hat{a_1}\textsuperscript{\textdagger}e^{-i\omega t}\nonumber\\
+i\chi^{(2)}(\hat{a_2}\textsuperscript{\textdagger}\hat{a_1}\hat{a_1}+\hat{a_1}\textsuperscript{\textdagger}\hat{a_1}\textsuperscript{\textdagger}\hat{a_2})
+m(\hat{a_1}\textsuperscript{\textdagger}\hat{a_3}+\hat{a_3}\textsuperscript{\textdagger}\hat{a_1})
\label{eq:1}
\end{eqnarray}
represented in the units of $\hbar$. The first term of the Hamiltonian reflects the energy of LSPR belongs to each oscillatory plasmon mode with $\hat{a_j}$($\hat{a_j}\textsuperscript{\textdagger}$), j=1-3 being the annihilation(creation) operators.  The second and last terms in Eq. (\ref{eq:1}) describe the linear and nonlinear interaction of driven and SH (LSPR) of AuNP with AgNR (dark mode) respectively where m accounts for the coupling strength between driven mode and dark mode and $\mathcal{K}$ is the coupling between SH and dark mode. Also, $\chi^{(2)}$ is defined as the strength of nonlinear mode derived from overlap integral \cite{Artvin2020}. The polarization of the driving field is along the x-axis as shown in Fig.\ref{fig1} (a) with frequency $\omega$. The equations of motion for $\alpha_1$, $\alpha_2$, and $\alpha_3$ amplitudes corresponding to driven plasmon mode (PM), second harmonic (SH) mode, and dark mode (DM) respectively are derived from the Heisenberg equation mentioned below.
\begin{eqnarray}
\partial_t\hat{a_j}=i[\hat{H},\hat{a_j}]
\label{eq:2}
\end{eqnarray}
bBy solving Eq.\ref{eq:2}, the driven-disscipative dynamics can be derived as: 
\begin{equation}
 \dot{\alpha_1}= -(i\Omega_1+\gamma_1)\alpha_{1}-i\chi^{(2)} \alpha_1^\ast\alpha_{2}-i m\alpha_{3}
 \label{eq:3}
\end{equation}
\begin{equation}
 \dot{\alpha_2}= -(i\Omega_2+\gamma_2)\alpha_{2}-i\chi^{(2)} \alpha_1^2 -i \mathcal{K}\alpha_{3}
 \label{eq:4}
\end{equation}
\begin{equation}
 \dot{\alpha_3}=-(i\Omega_3+\gamma_3)\alpha_{3}-i m\alpha_{1} -i \mathcal{K}\alpha_{2}
 \label{eq:5}
\end{equation}
Where the first terms in the Eq.(\ref{eq:3})- Eq.(\ref{eq:5}) are responsible for coherent dynamics with LSPR decay rates $\gamma_j$ associated with each mode. Our generalization of the model includes the coupling of the second harmonic mode with the long-lived dark mode (frequency of SH mode is very close to frequency of dark mode). As demonstrated in Fig.\ref{fig1} (a) this generalization constitutes a minimal model to examine the lifetime of SH mode and the effect of SH-DM induced enhancement in this process. In this context, no coupling of driven mode to dark mode is considered by putting m=0, and also this can be done by detuning dark mode with driven mode (these are at off-resonant conditions). To solve the nonlinear system of differential equations accounting for the linear and nonlinear plasmonic responses, we derive the solutions by solving the system through a numerical differential equation solver using the Runge-Kutta method retaining nonlinear dependence $\chi^{(2)}$ into account, in which we consider small (arbitrary) second-order nonlinear (overlap) factor $\chi^{(2)}=10^{-5}\omega$.\\
Also, we apply the initial approximation by ignoring the pump interaction term since we are interested in following the decay time from the peak of the field amplitude. At this point, the initial conditions of the linear and nonlinear responses of the plasmonic fields determine the driven-dissipative dynamics of the model system. After the application of excitation pulse the peak amplitude decays abruptly within tens of picoseconds, so coupling the plasmon amplitude in the peak state leads to the initial conditions, Eq.(\ref{eq:6}). 
\begin{equation}
\alpha_1(t)=1 \hspace{0.5cm}  \alpha_2(t)=1 \hspace{0.5cm}  
\alpha_3(t)=0.
\label{eq:6}
\end{equation}
After calculating the time-dependent solutions, we further examine the temporal dynamics of nonlinear responses induced by the incident driving field. In the following section, lifetime of non-linear plasmon mode is explained in detail before lifetime enhancement calculation.

\subsection{lifetime of nonlinear plasmon response}
\begin{figure}[hbt!]
\includegraphics[scale=0.45]{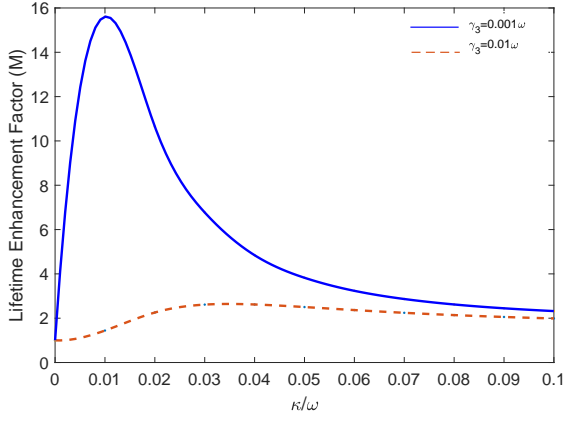}
\caption{\label{fig2} Lifetime enhancement factor calculated for the SH mode coupled to DM having different relaxation rates $\gamma_3=0.01\omega$ (dash line),$\gamma_3=0.001\omega$ (solid line) as a function of coupling strength $\mathcal{K}$.}
\end{figure}

\begin{figure}[hbt!]
\includegraphics[scale=0.45]{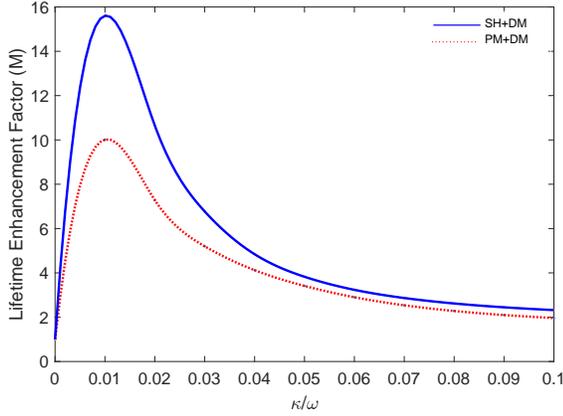}
\caption{\label{fig3} 
Lifetime enhancement factor calculated for second harmonic mode (SH) in two ways. 1) DM coupling to driven plasmon mode (PM) (dotted line), 2) DM coupling to SH mode (solid line), as a function of coupling strength $\mathcal{K}$.}
\end{figure}

 In a coupled state, the energy of the total system is in the form of a linear combination of driven and resonant oscillatory modes, $E_{j}= \sum_{j=1}^{3} |\alpha_{j}|^2$ which define the total number of plasmons in the coupled system. For a medium with dispersive dielectric function, the decay rate of each plasmon mode is given by \cite{Shahbazyan2016},
\begin{equation}
\Gamma_{j}=\frac{\mathcal{Q}_{j}}{\mathcal{U}_{j}}
\label{eq:7}
\end{equation}
where $\mathcal{U}_{j}$ represents the eigen energies of plasmon modes under consideration which in our case is the electric field intensity of second harmonic mode $\alpha_2$ coupled to dark mode $\alpha_3$. $Q_{j}$ is the dissipated power of the coupled system, together with the eigen energies, yields the lifetime of the induced coupled eigenmodes of plasmonic fields \cite{Yildiz2020},
\begin{equation}
\tau=\frac{\int_{0}^{\infty} t|E_{j}|^2}{\int_{0}^{\infty} |E_{j}|^2}.
\label{eq:8}
\end{equation}
In the weak coupling regime, the enhancement factor (EF), \textit{M}, is determined by taking the ratio of Eq.(\ref{eq:8}) containing $\mathcal{K}=0$ in the denominator for the uncoupled system as written in Eq.(\ref{eq:9}). Following Eq.(\ref{eq:9}) the nonlinear lifetime enhancement factor of SH mode coupled to DM is shown in the Fig.\ref{fig2} as a function of coupling strength $\mathcal{K}$. 
\begin{equation}
M=\frac{\tau[\mathcal{K}\ne0]}{\tau[\mathcal{K}=0]}
\label{eq:9}
\end{equation}
Here, all the rates/frequencies are taken in the units of $\omega$ (driving frequency). The relaxation rates and the plasmon frequencies are parameterized as follows $\gamma_{1}/\omega=0.1$, $\gamma_{2}/\omega=0.1$, $\Omega_{1}/\omega=1.0$, $\Omega_{2}/\omega=2.0$, and $\Omega_{3}/\omega=1.95$ for driven, second harmonic and dark mode respectively in the dimensions of excitation frequency $\omega=3.8 \times 10^{15}$ rad/sec. For these parameters, the critical coupling parameter $\mathcal{K}$ is ranged from 0 to $0.1\omega$. These parameters are similar to those adopted in simulations after replacing the dark mode from the quantum emitter for simplicity. Finally, we introduce two decay rates $\gamma_{3}/\omega$=0.01, 0.001 for the oscillating field of dark plasmon mode supported by AgNR interplaying with SH mode. Comparison of lifetime enhancement factors for two different damping rates is shown in Fig.\ref{fig2}.
The curves show the rising trend in the closed proximity of coupling regime where $\mathcal{K}$ is around $0.01\omega$ for the small decay rate of DM $(10^{-3}\omega)$ and it falls off smoothly as soon as the coupling is slightly increased to $0.02\omega-0.03\omega$. Whereas the enhancement factor for $\gamma_3=0.01\omega$ peaks around 3 at $\mathcal{K}=0.035\omega$. For the case of $\gamma_3=0.001\omega$, the lifetime of coupled SH is 13 times more enhanced than in the case of decay rate, $\gamma_3=0.01\omega$, showing the ultimate enhancement in the lifetime of nonlinear mode (SH) by coupling it to the dark mode. Moreover, when dark mode is coupled to the driven mode we analyze the enhancement in the lifetime of SH mode by comparing it to the system when SH mode is coupled to dark mode. Fig.\ref{fig3} shows the curves for the SH lifetime enhancement factor, where we obtained an increase in SH lifetime (solid curve) without modifying the first mode. This is because of coherent resonance of long-lived dark mode which enhances the SH efficiency and as a result, oscillating time of quickly decaying nonlinear plasmonic field exceeds numerously.
In the next section, we introduce the QE coupling with the SH mode of AuNP and solve the nonlinear system through a numerical differential equation solver using the Runge-Kutta method.  
\subsection{AuNP-QE Coupled Nonlinear Plasmonic System}
By replacing the eigen energy of the dark mode in Eq.(\ref{eq:1}) to the eigen energy of quantum emitter, the Hamiltonian of the coupled AuNP-QE system becomes as following.
\begin{align}
\hat{\mathcal{H}}= & \sum_{j=1}^{2}\omega_j\hat{a_j}\textsuperscript{\textdagger}\hat{a_j}+\omega_{q}\ket{e}\bra{e}+\mathcal{K}(\hat{a_2}\textsuperscript{\textdagger}\ket{g}\bra{e}+\hat{a_2}\ket{e}\bra{g})\nonumber \\  & +\varepsilon_1\hat{a_1}\textsuperscript{\textdagger}e^{-i\omega t}+i\chi^{(2)}(\hat{a_2}\textsuperscript{\textdagger}\hat{a_1}\hat{a_1}+\hat{a_1}\textsuperscript{\textdagger}\hat{a_1}\textsuperscript{\textdagger}\hat{a_2})
\label{eq:10}
\end{align}
After applying Eq.(\ref{eq:2}), we drive the equations of motion of the system, in which we are interested in temporal enhancement of the nonlinear plasmon mode $\alpha_2$.
\begin{equation}
 \dot{\alpha_1}= -(i\Omega_1+\gamma_1)\alpha_{1}-i\chi^{(2)} \alpha_1^\ast\alpha_{2}-i g\rho_{ge}
 \label{eq:11}
\end{equation}
\begin{equation}
 \dot{\alpha_2}= -(i\Omega_2+\gamma_2)\alpha_{2}-i\chi^{(2)} \alpha_1^2 -i \mathcal{K}\rho_{ge}
 \label{eq:12}
\end{equation}
\begin{equation}
 \dot{\rho_{ge}}=-(i\Omega_q+\gamma_q)\rho_{ge}+i (g\alpha_{1} + \mathcal{K}\alpha_{2})(\rho_{ee}-\rho_{gg})
 \label{eq:13}
\end{equation}
\begin{equation}
 \dot{\rho_{ee}}=-\gamma_{ee}\rho_{ee}+i (g\alpha_{1}^\ast + \mathcal{K}\alpha_{2}^\ast)\rho_{ge}
 \label{eq:14}
\end{equation}
After solving the nonlinear system of differential equations numerically, we perform the analysis by calculating the lifetime of the nonlinear plasmonic mode using Eq.(\ref{eq:8}) and related enhancement factor determined by Eq.(\ref{eq:9}) depending on the SH-QE coupling strength $\mathcal{K}$. Here, the frequencies and relaxation rates for PM and SH are the same as in the case of dark mode coupling except for the introduction of density matrix element $\rho_{ge}$ and resonant transition frequency of quantum emitter $\Omega_{q}/\omega=1.95$ with decay rate $\gamma_{q}/\omega=10^{-6}$, where the nonlinear susceptibility is taken as $\chi^{(2)}=10^{-5}\omega$. 
\begin{figure}[hbt!]
\includegraphics[scale=0.45]{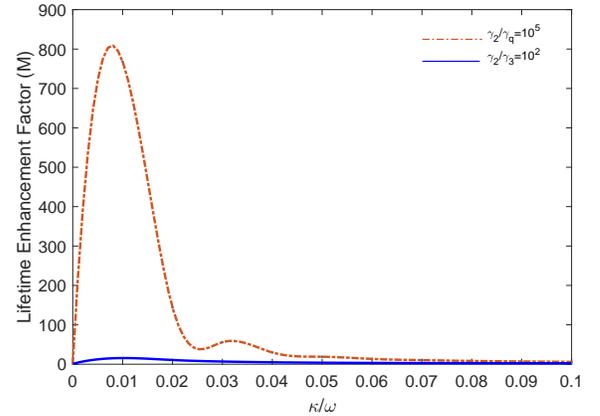}
\caption{\label{fig4} 
Lifetime enhancement spectra for the nonlinear second harmonic plasmonic mode compared for two cases. Dash-dotted: when the ratio of decay rates of SH to QE is $\gamma_2/\gamma_q=10^5$. Blue solid line: when decay rate ratio of SH to DM is $\gamma_2/\gamma_3=10^2$.}
\end{figure}
The lifetime enhancement factor determined from the nonlinear solutions follows the characteristic curve with ultrahigh enhancement than for the case of dark mode coupling. This huge enhancement accounts for the change in the decay rate ratio of SH to QE which is around $\gamma_2/\gamma_q=10^5$ in contrast with the decay ratio of SH mode to dark mode $\gamma_2/\gamma_3=10^2$. The SH lifetime enhancement factor is shown with dash-dotted line in Fig.\ref{fig4}, demonstrates maximum enhancement with a factor of 800 (compare peak values of the dash and line curves). The curves decay down smoothly by bringing QE close to AuNP, the strong coupling regime for hybridization of two modes.\\
After evaluating the lifetime as a ratio of average power radiated to total energy of LSPR, we introduce the electric field intensity of SH mode and DM as a function of time quantifying the efficiency of these modes in a coupled and uncoupled system. We also compare the plots of absolute values of SH efficiency with and without coupling to the quantum emitter. The Fig.\ref{fig5} and Fig.\ref{fig6} show the field intensities as a function of time. Without any coupling (solid line), the field intensity of SH mode linearly falls to zero around 26 fs of time which is the natural oscillating time of the plasmonic field. When the dark mode is coupled to the nonlinear system, the SH signal efficiency (dash line) in Fig.\ref{fig5} is slightly reduced.  
\begin{figure}[hbt!]
\includegraphics[scale=0.45]{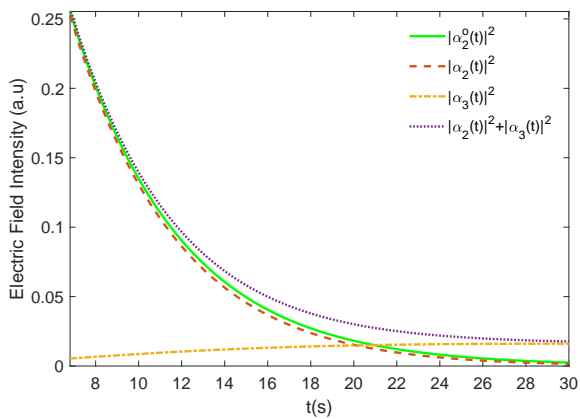}
\caption{\label{fig5} 
The electric field intensity is shown as a function of time. Dash and solid lines show data of SH field amplitude with and without coupling respectively at resonance frequency $\Omega_{2}=2.0\omega$ and relaxation $\gamma_{2}=0.1\omega$. The dash-dotted line is for the dark mode with $\Omega_{3}=1.95\omega$ and $\gamma_{3}=0.001\omega$. The red dotted line shows the result for the combined intensity of SH-DM at $\mathcal{K}=0.015\omega$.}
\end{figure}
This is primarily because AgNR does not generate SH directly and acts as an absorber at the fundamental frequency. However, when conditions for the polariton gain are met and both modes add up together, the picture changes drastically.
\begin{figure}[hbt!]
\includegraphics[scale=0.45]{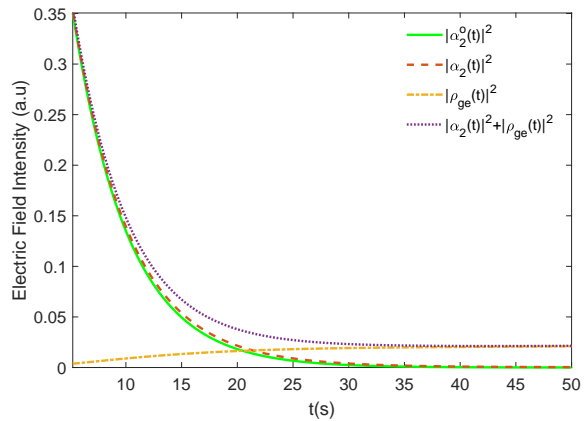}
\caption{\label{fig6} 
The electric field intensity is shown as a function of time. The red dashes and green solid lines show SH intensity with and without coupling respectively at resonance frequency $\Omega_{2}=2.0\omega$ and relaxation $\gamma_{2}=0.1\omega$. The curve with (-.) is for the quantum emitter with $\Omega_{q}=1.95\omega$ and $\gamma_{q}=10^{-6}\omega$. The blue dotted line shows the result for the combined intensity of SH-QE at $\mathcal{K}=0.015\omega$.}
\end{figure}
In the weak coupling threshold ($\mathcal{K}=0.015\omega$), the dark mode has a long oscillation time which, when couples to SH mode (solid line), extends its field intensity and longevity for about 30 fs more in the combined system (dotted curve) before relaxing to its natural eigenstate. Whereas when QE is coupled to nonlinear (SH) mode (dash line) in Fig.\ref{fig6}, field time extends slightly, and upon oscillating together with the QE (dotted line), the field elongates up to 50 fs which signifies the effect of coupled states on the lifetime of SH mode without modulating the driven mode. Now, to have a complete envision of temporal dynamics of our nonlinear plasmonic system, we realized a computational model in the next section and analyze the changes in lifetime enhancement when AuNP as an SH converter couple to quantum emitter resonantly.
\section{Computational Model}
Here, the nonlinear plasmonic system is analyzed using the finite difference time domain method (FDTD) to analyze the nonlinear temporal response of the system. The FDTD simulations are performed with a second harmonic converter facilitated by a 50 nm AuNP with nonlinear susceptibility $\chi^{(2)}=3.1 \times 10^{-5}$. We use experimental values rather than Drude Model for the dielectric function of Au \cite{Rakic:98}. 
\begin{figure*}[hbt!]
\includegraphics[scale=0.5]{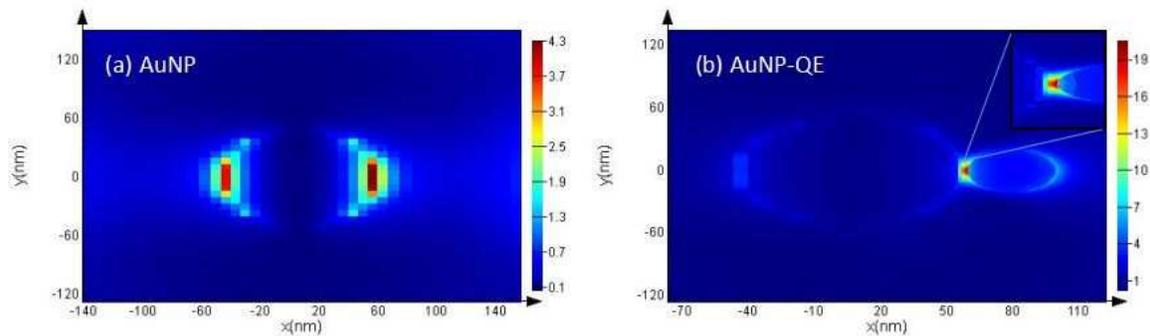}
\caption{\label{fig7} Electric field profile of LSPR: (a) Single AuNP 50nm size with second harmonic LSPR on each side excited by a plane wave. (b) Interaction of SH signal $2\omega$ supported by AuNP with QE at resonant transition frequency, the inset showing the intense hotspot in the nanogap.}
\end{figure*}
\subsection{Time domain numerical simulations of SH coupled to QE}
The spherical AuNP is placed under the action of a pump pulse with excitation wavelength range $\lambda_{exc}$=266 nm-270 nm correspond to SH frequency. The pulse excites SH field on the right and left poles of the particle, corresponding to the x-polarization of the pulse as illustrated in Fig.\ref{fig7} (a). The hotspot has a width of about 20 nm from 55 nm to 75 nm region as expected to be due to localized surface plasmon resonance(LSPR) at the outer surface of the nanoparticle. Now to scrutinize the spatial and temporal dynamics of the hotspot field, a quantum emitter of 20 nm size is placed close to the hotspot field with a distance around 10 nm from the AuNP. Similar to our nonlinear plasmonic model our computational model treats QE as a long lifetime quantum object. When QE is tuned in resonance ($\Omega_2=\omega_{eg}$) with SH mode, a strong hotspot emerges in between the AuNP and QE as seen in Fig.\ref{fig7} (b), the inset shows a more prominent view of near field coupling. Combined with a two-level quantum emitter the system exhibits an increase in SH plasmon mode efficiency as well as extension in oscillating time. We also note that the x-polarized SH signal efficiency enhanced significantly at the nanogap. We observe this effect predominantly for the linear plasmon mode. This is again due to the resonant interaction and the fact that emission by the QE is isotropic thus leading to the SH enhancement only in the direction perpendicular to the incident field polarization.  
To evaluate the plasmon response efficiency with and without quantum emitter, we calculate the power spectra separately using point monitor field-time response.
\subsection{Field-time response of nonlinear(SH) mode coupled to QE}
In this section, we evaluate the field-time response of nonlinear plasmon resonance upon interaction with quantum emitter having resonant transition frequency and nanosecond relaxation time. We place a point monitor at a 65 nm position close to the peak of the hotspot field. 
\begin{figure}[hbt!]
\includegraphics[scale=0.35]{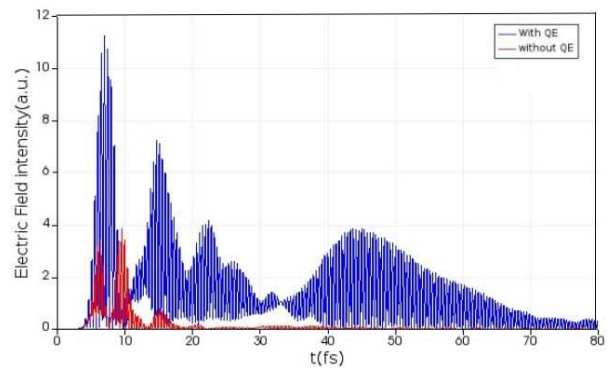}
\caption{\label{fig8} 
Comparison of SH field intensity as a function of oscillation time. The red line shows data without QE  Blue line shows field oscillation with QE when $\Omega_2=\omega_{eg}$.}
\end{figure}
In Fig.\ref{fig8} the electric field intensity is shown as a function of time. In the absence of QE, the power spectrum (red line) exhibits a maximum peak amplitude around 3 indicating response (LSPR) of AuNP and it decays at 20 fs with no further oscillations. Bringing the QE in the vicinity of the LSPR hotspot and setting the time monitor in between the particles and field monitor around the system, we observed an intense hotspot field at 65 nm in the region between QE and AuNP. In this regime, the system acquires gain (i.e., amplification) thrice the intensity of driven mode resulting from the polariton-plasmon coupling through the absorption at the QE transition frequency.
Interestingly, due to strong coupling between QE and LSPR, the amplification is still showing up after 20 fs nearby plasmon resonance and it continues to oscillate till 80 fs. As the decay time of the quantum emitter is around a nanosecond so it starts oscillating with the plasmonic field. Having QE transition frequency close to plasmon mode frequency provides the system coherent resonance state in which the amplitude of the field amplifies by superposition and the system acquires enough energy to oscillate for an extended time.\\
It is important to emphasize the difference between theoretical models and three-dimensional numerical simulations. Firstly, the numerical simulations take into account multiple channels that lead to losses thus noticeably diminishing the SH efficiency at some point around 35 fs as we can see in the case of QE coupling to nanoparticle Fig.\ref{fig8}. Such channels include radiation to the far-field and spatially dependent local field distribution, which both significantly lower the coupling between QE and corresponding resonant plasmon mode. Secondly, the effective second-order susceptibility, the parameter used directly in the simulations, most like depends on the geometry of the particle which affects the efficiency of SH as well.
\section{conclusion}
We have investigated the temporal dynamics of nonlinear plasmonic systems. Theoretical and computational models are optimized, consisting of nanostructures interacting with quantum objects supporting longer lifetime characteristics as compare to shortly lived plasmon resonances, under coherent pumping. We demonstrated that in the weak coupling regime the combinations of differently persisting plasmonic systems exhibit noticeable enhancement in the lifetime of the second harmonic plasmonic field. The harmonic oscillator model accounting for the LSPR complex amplitudes of each mode provides us with an insight into the mechanism behind the generation of linear and nonlinear plasmon modes and their interplay with the dark mode or QE longevity.  
Only a weak interaction elongates the time of a rapid decaying oscillating field and improves efficiency. For a more realistic visualization, we have performed the numerical simulation of a field-time profile of a plasmonic system by coupling AuNP to QE. The simulation supports the idea of amplifying the oscillating time of nonlinear(SH) LSPR together with the resonant QE without disturbing the lifetime of fundamental plasmon mode. Our idea can be used for distinguishing linear and non-linear plasmonic responses for a variety of applications. 
\appendix
\nocite{*}


%

\end{document}